
%
%
%
%
\def\today{\ifcase\month\or January\or February\or March\or April\or May\or
June\or July\or August\or September\or October\or November\or December\fi
\space\number\day, \number\year}
%
%
\newcount\notenumber

\def\note{\global\advance\notenumber by 1 \footnote{$^{\the\notenumber}$}}
%
%
\newif\ifsectionnumbering
\newcount\eqnumber
\def\cleareqnumber{\eqnumber=0}
\def\numbereq{\global\advance\eqnumber by 1
\ifsectionnumbering\eqno(\the\secnumber.\the\eqnumber)\else\eqno(\the\eqnumber)
\fi}
\def\eqalinno{{\global\advance\eqnumber by 1}
\ifsectionnumbering(\the\secnumber.\the\eqnumber)\else(\the\eqnumber)\fi}
\def\name#1{\ifsectionnumbering\xdef#1{\the\secnumber.\the\eqnumber}
\else\xdef#1{\the\eqnumber}\fi}

\sectionnumberingtrue
%
%
\newcount\refnumber

\immediate\openout1=refs.tex
\immediate\write1{\noexpand\frenchspacing}
\immediate\write1{\parskip=0pt}
\def\ref#1#2{\global\advance\refnumber by 1%
[\the\refnumber]\xdef#1{\the\refnumber}%
\immediate\write1{\noexpand\item{[#1]}#2}}
\def\tie{\noexpand~}

%
%
\font\twelvebf=cmbx10 scaled \magstep1
\newcount\secnumber

\def\newsection#1.{\ifsectionnumbering\cleareqnumber\else\fi%
	\global\advance\secnumber by 1%
	\bigbreak\bigskip\par%
	\line{\twelvebf \the\secnumber. #1.\hfil}\nobreak\medskip\par}
%
%
%
\def \sqr#1#2{{\vcenter{\vbox{\hrule height.#2pt
	\hbox{\vrule width.#2pt height#1pt \kern#1pt
		\vrule width.#2pt}
		\hrule height.#2pt}}}}

%
%
%
\newdimen\fullhsize
\def\fiddle{\fullhsize=6.5truein \hsize=3.2truein}
\def\fullline{\hbox to\fullhsize}
\def\mkhdline{\vbox to 0pt{\vskip-22.5pt
	\fullline{\vbox to8.5pt{}\the\headline}\vss}\nointerlineskip}
\def\mkftline{\baselineskip=24pt\fullline{\the\footline}}
\let\lr=L \newbox\leftcolumn
\def\twocolumns{\fiddle
	\output={\if L\lr \global\setbox\leftcolumn=\columnbox
		\global\let\lr=R \else \doubleformat \global\let\lr=L\fi
		\ifnum\outputpenalty>-20000 \else\dosupereject\fi}}
\def\doubleformat{\shipout\vbox{\mkhdline
		\fullline{\box\leftcolumn\hfil\columnbox}
		\mkftline} \advancepageno}
\def\columnbox{\leftline{\pagebody}}
\def\pr#1 {Phys. Rev. {\bf D#1\tie }}
\def\pl#1 {Phys. Lett. {\bf #1B\tie }}
\def\prl#1 {Phys. Rev. Lett. {\bf #1\tie }}
\def\np#1 {Nucl. Phys. {\bf B#1\tie }}
\def\ap#1 {Ann. Phys. (NY) {\bf #1\tie }}
\def\cmp#1 {Commun. Math. Phys. {\bf #1\tie }}
\def\imp#1 {Int. Jour. Mod. Phys. {\bf A#1\tie }}
\def\tie{\noexpand~}
\def\ov{\overline}
\def\s{(\sigma)}
\def\se{(\sigma+\epsilon)}
\def\sp{(\sigma ')}
\def\sep{(\sigma'+\epsilon')}
\def\spme{(\sigma'-\epsilon)}
\def\spaeaep{(\sigma'+\epsilon+\epsilon')}
\def\spaepme{(\sigma'+\epsilon'-\epsilon)}
\def\Tb{\overline T}
\def\d{\delta(\sigma-\sigma ')}
\def\dpr{\delta'(\sigma-\sigma ')}

\def\dpppr{\delta'''(\sigma-\sigma ')}
\def\de{\delta(\sigma-\sigma ' + \epsilon)}
\def\dep{\delta(\sigma-\sigma ' - \epsilon ')}
\def\deep{\delta(\sigma-\sigma ' + \epsilon - \epsilon')}
\def\dpreep{\delta'(\sigma-\sigma ' + \epsilon -
\epsilon')}
\def\ints{\int d\sigma\,}

\def\dx{\partial X}
\def\dbx{{\overline\partial}X}
\def\dsx{\partial^2X}
\def\dbsx{{\overline\partial}^2X}
\parskip=15pt plus 4pt minus 3pt
\headline{\ifnum \pageno>1\it\hfil  An
Infinite-Dimensional Symmetry
	$\ldots$\else \hfil\fi}
\font\title=cmbx10 scaled\magstep1
\font\tit=cmti10 scaled\magstep1
\footline{\ifnum \pageno>1 \hfil \folio \hfil \else
\hfil\fi}
\raggedbottom
\rightline{\vbox{\hbox{RU93-8}\hbox{DOE/ER40325-2TASKB}
\hbox{CTP-TAMU-2/94}\hbox{CERN-TH.7022/93}\hbox{ACT-29/94}}}
\vfill
\centerline{\title AN INFINITE-DIMENSIONAL SYMMETRY
ALGEBRA}
\vskip 20pt
\centerline{\title IN STRING THEORY}
\vfill
\centerline{\title Mark Evans$^{(a)}$,
Ioannis Giannakis$^{(b),(c)}$
and D.~V.~Nanopoulos$^{(b), (c), (d)}$}
\medskip
\centerline{$^{(a)}${\tit Physics Department, Rockefeller
University}}
\centerline{\tit 1230 York Avenue, New York, NY
10021-6399}
\medskip
\centerline{$^{(b)}${\tit Center for Theoretical Physics,
Texas A{\&}M
University}}
\centerline{\tit College Station, TX 77843-4242}
\medskip
\centerline{$^{(c)}${\tit Astroparticle Physics Group,
Houston Advanced
Research Center (HARC)}}
\centerline{\tit The Mitchell Campus, Woodlands, TX 77381}
\medskip
\centerline{$^{(d)}${\tit Theory Division, CERN}}
\centerline{\tit CH-1211 Geneva 23, Switzerland}
\vfill
\centerline{\title Abstract}
\bigskip
{\narrower\narrower
Symmetry transformations of the space-time fields of
string theory are
generated by certain similarity transformations of the
stress-tensor
of the associated conformal field theories.
This observation is complicated by the fact that, as we
explain, many of the operators we habitually use in string
theory (such
as vertices and currents) have ill-defined commutators.
However, we
identify an infinite-dimensional subalgebra whose
commutators are not
singular, and explicitly calculate its structure
constants. This
constitutes a subalgebra of the gauge symmetry of string
theory, although
it may act on auxiliary as well as propagating fields. We
term this
object a {\it weighted tensor algebra}, and, while it
appears to be a
distant cousin of the $W$-algebras, it has not, to our
knowledge, appeared
in the literature before.\par}
\vfill\vfill\break

\newsection Introduction.

Closed string theory is the most promising candidate for a
complete dynamics
of elementary particles, yet its usual formulation as a
first quantized
theory is widely regarded as being rather unsatisfactory.
First quantized
string theory yields rules for the calculation of S-matrix
elements in
particular backgrounds, but obscures a more global ({\it
i.e.~}background
independent) view of the theory, and shrouds in mystery
its deeper
principles. Symmetry, that most powerful tool of the
theoretical
physicist, must surely be counted among these principles.

In this paper we shall describe an infinite-dimensional
subalgebra of
the full symmetry algebra of a $D$-dimensional string
theory. We shall also point out that many of the operators
we habitually use in string theory are not well-defined---their
commutators are singular.

There are several approaches to the study
of the symmetries of string theory in the literature;
let us briefly survey those of which we are aware.
One natural approach
is to attempt to formulate an action for a field
theory and then study
its invariances. Constructing a string field theory
has proved a formidable task for reasons not obviously associated
with the problem of symmetry. Nevertheless considerable
progress has been made recently in this direction
(see, for example,
\ref{\dfes}{B. Zwiebach \np390 (1993), 33, hep-th/9206084;
A. Sen and B. Zwiebach preprint MIT-CTP-2244, hep-th/9311009.}, and
references therein to the earlier literature),
although we think it fair to say that a manifestly
background-independent
formulation is still missing.

Other work has avoided the difficulties
associated with going off-shell by working in the
first-quantized formalism. The most physically direct approach
is the {\it gedanken\/}-phenomenology of Gross, Mende and Manes
\ref\gross{D.\tie Gross, \prl60 (1988), 1229; D.\tie Gross and
P.\tie Mende, \np303 (1988), 407; D.\tie Gross and J.\tie Manes,
\np326 (1989), 73.}, who derived expressions for the asymptotic form of
scattering amplitudes at high energy. For our purposes, the important
lesson from this work is that this asymptotic form is independent
of the states on the external legs. This strongly suggests the
existence of an enormous, spontaneously broken symmetry, but sheds
little light on its actual form. Isberg, Lindstrom and Sundborg
had the related idea of studying symmetries in the limit of
zero tension \ref\lindstrom{J.\tie Isberg, U.\tie Lindstrom
and B.\tie Sundborg, \pl293 (1992), 321, hep-th/9207005.}, which
should correspond to the same high energy regime.

A second approach \ref{\genfn}{G.\tie Veneziano,
\pl167 (1986), 388; J.\tie Maharana and G.\tie Veneziano,
\pl169 (1986), 177;
\np283 (1987), 126; A.\tie Polyakov, Physica
Scripta {\bf T15} (1987), 191;
C.\tie Hull and E.\tie Witten, \pl160 (1985), 398; C.\tie Hull and
P.\tie Townsend, \pl178 (1986), 187.},
\ref{\ovre}{M.\tie Evans and B.\tie Ovrut, \pl214 (1988), 177;
\pr39 (1989), 3016; M.\tie Evans, J.\tie Louis and B.\tie Ovrut,
\pr35 (1987), 3045.},
treats the partition function of the non-linear
sigma model (somewhat formally) as a generating functional for string
amplitudes. This partition function is, of course, a functional integral,
and, ``changes of variables," in this functional integral generate
invariances which, in turn, translate into Ward identities among
amplitudes. This approach has the advantage of yielding directly the
transformations on space-time fields, but (in our view) is hampered by
the difficulties involved in defining the relevant functional measures.

The third type of approach focuses on algebraic aspects of string
theory. It has long been recognized \ref{\band}{T. Banks and L. Dixon
\np307 (1988), 93.} that conserved currents (``current," meaning a (1,0)
or (0,1) primary field) generate unbroken gauge symmetries. This notion
was generalized to include currents that are conserved anywhere in the
deformation class of a CFT \ref\giveon{A.\tie Giveon and M.\tie Porrati,
\np355 (1991), 422; A.\tie Giveon and A.\tie Shapere, \np386 (1992), 43,
hep-th/9203008.}. However, this is still a special case of the method we
shall employ in this paper; symmetries (broken and unbroken) are generated
by (inner) automorphisms of the operator algebra of a deformation
class of CFT's \ref{\evao}{M. Evans and B. Ovrut \pr41 (1990), 3149;
\pl231 (1989), 80.}, \ref{\egw}{M. Evans and I. Giannakis,
\pr44 (1991), 2467.}.
Moore has recently used this method (in the
Batalin-Vilkovisky formalism) to derive an infinite set of relations
that were sufficient to determine all amplitudes with fewer than
twenty-six external legs \ref\moore{G.\tie Moore, Yale preprint
YCTP-P19-93, hep-th/9310026.}.
Physical consequences of the symmetries uncovered in this way have
also been
explored in \ref\emn{J.\tie Ellis, N.\tie Mavromatos and D.\tie
Nanopoulos, \pl288 (1992), 23, hep-th/9205107; \pl284 (1992), 43,
hep-th/9203012; \pl278 (1992), 246, hep-th/9112062; \pl272 (1991),
261, hep-th/9109027.}.

This method will be reviewed in more detail
in section 2. In section 3 we shall compute the
symmetry subalgebra
that is the main result of this paper, while in section 4
we shall explain
why computing the {\it full\/} symmetry algebra is
problematic. Indeed,
section 4 contains a message of independent significance:
operators that
play important r\^oles in the theory (such as vertices
and
currents) {\it do not actually exist}, as presently
defined, because
their commutators are sick. We shall argue that this
problem is not simply
a technical detail to be corrected, but is probably
insoluble. We find this
problem troubling, but its full significance is obscure to
us.

We end this introduction with a brief description of our
infinite-dimensional symmetry subalgebra. As has been argued
elsewhere [\evao], it is a {\it supersymmetry}, in that its
generators do not commute with the generators of Lorentz transformations and
mix excitations of different spin; it is spontaneously broken in flat
space-time, because not all generators commute with the stress tensor
of the free scalar CFT, and it transforms excitations of differing mass
into one another.

The algebra may be described abstractly in terms of the
following objects:
consider the set of all
covariant tensors of a $D$-dimensional vector space, paired
with a $k$-tuple of
positive-definite integer weights ($k$ is the rank of the
tensor,
and each weight is associated with an index on the
tensor). Introduce the
operators $\Delta_r$, where $r\leq k$ is a positive
definite integer, defined
by
$$
\Delta_r \left\{\phi^{(k)}, w_i\right\} =
	\sum_{l=1}^r \left\{\phi^{(k)},
(w_i+\delta_{il})\right\}.
		\numbereq\name{\eqdefdelta}
$$
Then elements of the algebra consist of the pairs
$\left\{\phi^{(k)}, w_i\right\}$, modulo the relations
$$
\eqalignno{
\lambda\left\{\phi^{(k)}, w_i\right\} + \mu\left\{\psi^{(k)},
	w_i\right\}
	- \left\{\lambda\phi^{(k)}+\mu\psi^{(k)}, w_i\right\}
	&\cong 0 &\eqalinno\name{\eqlinear}\cr
\Delta_k \left\{\phi^{(k)}, w_i\right\}
	&\cong 0 &\eqalinno\name{\eqtotder}\cr
\left\{\phi^{(k)}_{\ldots\mu\ldots\nu\ldots},
	(\ldots,w_m,\ldots,w_n,\ldots)\right\} -
	\left\{\phi^{(k)}_{\ldots\nu\ldots\mu\ldots},
	(\ldots,w_n,\ldots,w_m,\ldots)\right\} &\cong 0
	&\eqalinno\name{\equnordered}\cr}
$$
Equation (\eqlinear) says that, for fixed weights, tensor
addition is
vector space addition, but note that there is no such
property for the
weights.
Equation (\eqtotder) asserts that, given a basis element,
incrementing
each weight by one and summing yields an element that
should be identified
with zero.
Equation (\equnordered) identifies two basis elements if
their tensors are
transposes on a pair of indices and the corresponding
weights are interchanged,
{\it e.g.~}$\left\{\phi^{(2)}, (w_1,w_2)\right\}\cong
\left\{\phi^{(2)T}, (w_2,w_1)\right\}$ where
$\phi^{(2)T}_{\mu\nu}=
\phi^{(2)}_{\nu\mu}$.

The commutator of two such generators is
$$
\left[\left\{\psi_{\mu_{\cal S}},w_{\cal S}\right\},
\left\{\chi_{\nu_{\cal T}},v_{\cal T}\right\}\right] =
\sum_{\scriptstyle\cal U\subseteq S\atop\scriptstyle
\mathstrut P\colon
U\hookrightarrow T} {\cal C_{U,P}} \Delta_{\left|\cal
S-U\right|}^{\left(-1+
\sum_{i\in\cal U} w_i+v_{{\cal P}(i)}\right)}
\left\{\psi_{\mu_{\cal S}}\chi_{\nu_{\cal T}}
\prod_{i\in\cal U} \eta^{\mu_i
\nu_{{\cal P}(i)}},\; w_{\cal S-U}\oplus v_{\cal
T-P\left(U\right)}\right\}.
\numbereq\name{\eqcommutator}
$$
The indices on the tensors $\psi$ and $\chi$ are
themselves labeled by
index sets $\cal S$ and $\cal T$, {\it e.g.}
${1,\dots,k}$.
The sum is over all pairs consisting
of subsets
$\cal U\subseteq S$ and injections $\cal P\colon
S\hookrightarrow T$ (or,
equivalently, all pairs of subsets of $\cal S$ and $\cal
T$ of the
same size). $\cal S-U$ denotes the complement of $\cal U$
in $\cal S$
and $\left|\cal S\right|$ denotes the number of elements
in $\cal S$.
$\eta$ is the Minkowski metric of space-time.
The weights, $w_{\cal S-U}\oplus v_{\cal
T-P\left(U\right)}$,
are the weights of $\psi$ that do not correspond to
indices in $\cal U$
together with the weights of $\chi$ that do not correspond
to the indices
in the image of $\cal U$ under $\cal P$. The superscript
on $\Delta$
is a power. The coefficients are given by
$$
{\cal C_{U,P}} = {\prod_{k\in\cal
U}(-1)^{w_k}(w_k+v_{{\cal P}(k)}-1)!\over
	\left(\left(\sum_{k\in\cal U} w_k+v_{{\cal
P}(k)}\right)-1\right)!}
\numbereq\name{\eqstructconst}
$$

The formal statement of this commutator is thus rather
cumbersome and
intimidating, but the basic idea is simply that of Wick's
theorem. The sum over
$\cal U$ and $\cal P$ is a sum over all possible
contractions of the
tensors $\psi$ and $\chi$. The weights are those of the
uncontracted
indices, except that the weights of $\psi$ are augmented
in a suitable
way by the weights of the contracted indices through a power
of the operator $\Delta_{\cal |S-U|}$. The
similarity to Wick's theorem is, of course, no accident.

To the best of our (limited) knowledge, algebras of this type
have not appeared in the literature until now; we suggest that they
be termed {\it weighted tensor algebras}.

\newsection Deformations of Conformal Field Theories and
Symmetries of
String Theory.

In this section we shall review earlier work [\evao], [\egw]
on deformations of
conformal field theories and symmetries of string theory.
For more details
the reader is referred to the original papers, or the review
contained in \ref\dgmtalk{M.\tie Evans and I.\tie Giannakis in
S.\tie Catto and A.\tie Rocha, eds., {\it Proceedings of the
XXth International Conference on
Differential Geometric Methods in Theoretical Physics}, World
Scientific, Singapore (1992), hep-th/9109055.}.

We shall study the question of string
symmetries by finding transformations of the space-time
degrees of freedom that map
one solution of the classical equations of motion to
another that is physically
equivalent. Since, ``Solutions of the classical equations
of motion," are,
for the case of string theory \ref{\soln}{C.\tie Lovelace,
\pl135 (1984), 75; C.\tie Callan, D.\tie Friedan,
E.\tie Martinec and M.\tie Perry, \np262 (1985), 593; A.\tie Sen,
\pr32 (1985), 2102.}, two-dimensional conformal
field theories
\ref{\bpza}{A. Belavin, A. Polyakov and A. Zamolodchikov
\np241 (1984), 333; D. Friedan, E. Martinec and S. Shenker
\np271 (1986), 93.},
we are thus interested in physically equivalent conformal
field theories.

Any quantum mechanical theory (including a CFT) is defined
by an algebra
of observables, $\cal A$ (determined by the degrees of
freedom of the theory
and their equal-time commutation relations), a
representation of that algebra
and a distinguished element of $\cal A$ that generates
temporal evolution
(the Hamiltonian). (Note that for the same $\cal A$ we may
have many choices
of Hamiltonian, so that $\cal A$ may more properly be
associated with a
{\it deformation class\/} of theories than with one
particular theory.)
For a CFT, we further want $\cal A$ to be generated by
local fields,
$\Phi(\sigma)$ (operator valued distributions on a circle
parameterized
by $\sigma$), and we require not just a single
distinguished operator, but
two distinguished fields, $T(\sigma)$ and $\ov T(\sigma)$,
in terms of which
the Hamiltonian, $H$, and generator of translations, $P$,
may be written
$$
\eqalignno{H&=\int d\sigma (T\s +
\Tb\s)&\eqalinno\name{\eqham}\cr
P&=\int d\sigma (T\s - \Tb\s)&\eqalinno\name{\eqmom}\cr}
$$
and that satisfy Virasoro$\times$Virasoro:
$$
\eqalignno{[T\s, T\sp]&={- i c \over 24\pi}\dpppr
+2 i T\sp\dpr -  i T'\sp\d&{\global\advance\eqnumber by 1}
(\the\secnumber.\the\eqnumber a)\name{\eqvir}\cr
[\Tb\s,\Tb\sp]&={ i c \over 24\pi}\dpppr
-2 i\Tb\sp\dpr +  i\Tb'\sp\d&(\the\secnumber.\the\eqnumber
b)\cr
[T\s,\Tb\sp]&=0.&(\the\secnumber.\the\eqnumber c)\cr}
$$
Except on $\sigma$, a prime denotes differentiation.
$T$ and $\ov T$ are the non-vanishing components of the
stress tensor,
and must satisfy (\eqvir) if they are to generate
conformal transformations.
Also of interest are the so-called
{\it primary fields\/} of dimension $(d,\ov d)$, $\Phi\s$,
defined
by the conditions
$$
\eqalign{[T\s,\Phi_{(d,\overline d)}\sp]&=i d
\Phi_{(d,\overline d)}\sp
\dpr-( i/\sqrt2)\partial\Phi_{(d,\overline d)}\sp\d\cr
[\Tb\s,\Phi_{(d,\overline d)}\sp]&=- i\overline d
\Phi_{(d,\overline d)}
\sp\dpr-( i/\sqrt2)\overline\partial\Phi_{(d,\overline
d)}\sp\d\cr}
\numbereq\name{\eqprim}
$$

Clearly, then, two CFTs will be physically equivalent if
there is an isomorphism
between the corresponding algebras of observables, $\cal
A$, that maps
stress tensor to stress tensor. (The mapping of primary to
primary is then
automatic).
The simplest example of such an isomorphism
is an inner automorphism, or similarity transformation:
$$
\Phi(\sigma)\mapsto
e^{ih}\Phi(\sigma)e^{-ih}\numbereq\name{\eqinaut}
$$
for any fixed operator $h$. Thus the physics will be
unchanged if we change
a CFT's stress tensor by just such a similarity
transformation.

Now, the stress tensor is parameterized by the space-time
fields of the
string. For example,
$$
T(\sigma) = \textstyle {1\over2} G_{\mu\nu}(X)\partial
X^\mu\partial X^\nu
\numbereq\name{\eqgravback}
$$
corresponds to the space-time metric $G_{\mu\nu}$, with
all other fields
vanishing. Thus a similarity transformation (\eqinaut)
applied to $T$
will, in general, produce a change in $T$ which
corresponds to a change
in the space-time fields, {\it without changing the
physics}. This change
in the space-time fields is therefore a {\it symmetry\/}
transformation.
In this way, one may exhibit symmetries both familiar
(general coordinate
and two-form gauge transformations, regular non-abelian
gauge
transformations---including the Green-Schwarz
modification---for
the heterotic string [\ovre]) and unfamiliar (an infinite
class of spontaneously
broken, level-mixing gauge symmetries).

We may clarify
the way in which the change in the stress tensor may be
interpreted as a
change in the space-time fields by first considering the
more
general problem of deforming a conformal field theory (we
now consider
deformations which, while they preserve conformal
invariance,
{\it need not be symmetries\/}). It is straightforward to
show that, to first order, the Virasoro algebras (\eqvir)
are preserved by
deforming the choice of stress tensor by a so-called {\it
canonical
deformation},
$$
\delta T\s = \delta\Tb\s =
\Phi_{(1,1)}\s \numbereq\name{\eqcan}
$$
where $\Phi_{(1,1)}\s$ is a primary
field of dimension (1,1) with respect to the stress
tensor. We reiterate:
(\eqcan) does {\it not\/} in general correspond to a
symmetry transformation,
although it preserves conformal invariance.
Since (1,1) primary
fields are vertex operators for physical states, they are
in natural
correspondence with the space-time fields, and equation
(\eqcan) makes the
connection between changes of the stress tensor and
changes of the space-time
fields more transparent. Returning now to the problem of
symmetries, if we take
the generator $h$ in equation (\eqinaut) to be the zero
mode of an
infinitesimal (1,0) or (0,1) primary field (a current),
then it is
straightforward to see
that its action on the stress tensor is necessarily
a canonical deformation, as in equation (\eqcan),
which may easily be translated into a change in the
space-time fields. It is
well known that {\it conserved\/} currents generate
symmetries [\band], but
within the formalism described here, conservation is {\it
not\/}
necessary, a fact that does not seem to have been widely
appreciated.
Indeed, it may be shown that a non-conserved current
generates
a symmetry that is spontaneously broken by the particular
background being
considered [\evao].

Actually, we do not even need the generator, $h$, to be a
current. By
considering a few examples, it is easy to see that the
canonical deformation of equation (\eqcan) corresponds to
turning on
space-time fields in
a particular gauge (something like Landau or harmonic
gauge), and so
symmetries generated by zero-modes of currents preserve
this gauge
condition, since they generate canonical deformations.
Furthermore, the commutator of zero modes of currents is
not necessarily itself the zero-mode of a current. Thus
restricting
the generators $h$ in this way we are dealing with a
subset of the symmetry
generators that do not even form a subalgebra.

On the other hand, equation (\eqcan) is {\it not\/} the
most general infinitesimal deformation that preserves the
Virasoro
algebras (\eqvir). In [\egw] we showed that, for the
massless degrees
of freedom of the bosonic string in flat space, we could
find a distinct
deformation of the stress tensor for each solution of the
linearized
Brans-Dicke equations. This correspondence was found by
considering the
general translation invariant {\it ansatz\/} of naive
dimension two for
$\delta T$;
$$
\delta T=H^{\nu\lambda}(X)\dx_\nu\dbx_\lambda+
A^{\nu\lambda}(X)\dx_\nu\dx_\lambda+B^{\nu\lambda}(X)
\dbx_\nu\dbx_\lambda
+C^\nu(X)\dsx_\nu+D^\lambda(X)\dbsx_\lambda,\numbereq\name{
\eqansatz}
$$
with a similar, totally independent {\it ansatz\/} for
$\delta\Tb$.
The fields
$H^{\mu\nu}$ {\it etc.}~turn out to be characterized in
terms of solutions
to the linearized Brans-Dicke equation when we demand that
the deformation
preserves (to first order) the Virasoro algebras (\eqvir).
By considering
this more general {\it ansatz}, we get more than just
covariant equations of
motion---we may also understand a larger set of symmetry
generators, $h$.
Indeed, any generator that preserves the form of the {\it
ansatz\/}
(\eqansatz)
must necessarily generate a change in the stress tensor
that corresponds
to a change in the space-time fields.

The condition that $\delta T$ be of naive dimension two
(with which we
shall soon dispense) is preserved if $h$ is of naive
dimension zero. The
condition of translation invariance is
$$
\left[P,\delta T\s\right] = -i\delta T'\s,
\numbereq\name{\eqtransinv}
$$
which may be preserved by demanding that $h$ commute with
$P$, the
generator of translations, (\eqmom). (Equation (\eqtransinv)
may also be thought
of as a gauge condition, but not one that has any obvious
interpretation
in terms of the space-time fields). Taken together, these
conditions
restrict $h$ to be the zero-mode of a field of naive
dimension one.

The lesson to be drawn from this massless example is
clear: the way to
introduce space-time fields unconstrained by gauge
conditions is to consider
an {\it arbitrary translation invariant ansatz\/} for the
deformation
of the stress tensor, and to ask only that it preserve the
Virasoro
algebras. To move beyond the massless level, we simply
drop the requirement
that the naive dimension be two. We argued in [\egw] that
this was likely to
introduce auxiliary fields beyond the massless level, as
with the superfield
formulation of supersymmetric theories, but so be it.
(Indeed, this
whole formulation
of string theory is rather akin to a superspace approach,
with
$T$ and $\Tb$ as superfields and
derivatives of the world-sheet scalars playing the r\^ole
of the
odd coordinates of superspace).

Having dropped any
requirement on the naive dimension of $\delta T$, we know
that {\it any}
operator $h$ that commutes with $P$ will generate a
symmetry transformation
on our space-time fields (possibly including the
auxiliaries). Furthermore,
the centralizer of $P$ is necessarily an algebra---the
commutator of the
zero modes of two fields is itself a zero-mode. We are
therefore in a
position to ask questions about the symmetry algebra that
were beyond us
as long as we restricted our generators to be zero-modes
of currents.

As we shall explain in section 4, there are severe
obstacles to calculating
the full centralizer of $P$, but in the next section we
shall consider
a tractable subalgebra. Any $D$-dimensional string theory
has a formulation
of its CFT in terms of at least $D$ scalar fields, $X$,
(among others).
The algebra we shall consider is generated by zero modes
of fields
that do not contain functions of the $X$ themselves, but
rather only holomorphic derivatives thereof.

\newsection Symmetry Algebra.

In the last section, we explained why any
zero-mode in the operator algebra of
a deformation class of CFTs generates
a symmetry transformation of string
theory. The symmetry algebra is just the
algebra of these zero-modes. In this
section we shall explicitly calculate the
structure constants of an
infinite-dimensional subalgebra of this full
symmetry algebra---that generated
by arbitrary normal-ordered products of
holomorphic derivatives of the free
scalar field, $X$, that is, operators of the form:
$$
h= \ints {\phi_{\mu\nu\cdots\rho}}{\partial^{w_1}}X^{\mu}{\partial^{w_2}}
X^{\nu} \cdots {\partial^{w_n}}X^{\rho}. \numbereq\name\eqkorad
$$
As such, it differs from other infinite-dimensional algebras that have
appeared
in string theory (Virasoro, affine, $W$), all of which are
infinite because of
the infinite moding of a (usually) finite set of fields.
By contrast, our
algebra arises from the zero-modes of an infinite set of fields.

Here and throughout the paper, the light-cone ``derivatives" $\partial$ and
$\ov\partial$ are to be interpreted, ``as if in free-field theory." That is,
$$
\eqalignno{
\dx^\mu\s &= \left(\eta^{\mu\nu}\pi_\nu\s + {X^\mu}'\s\right)/\sqrt2,
&\eqalinno \name\eqdefdx\cr
\dsx^\mu\s &= \left(\eta^{\mu\nu}{\pi_\nu}'\s + {X^\mu}''\s\right),\qquad
\hbox{etc.}, &\eqalinno \name\eqdefdsx\cr
}
$$
the symbol on the left being simply
a shorthand notation for the operator on
the right, which should make sense for
{\it all\/} the CFTs in the deformation
class, even though the object in equation
(\eqdefdx), for example, is only the
light-cone derivative of the scalar field
$X^\mu$ for the particular case of
free field theory.

Operators of the form of equation (\eqkorad)
are clearly specified by a tensor
on $R^D$, $\phi$, and a weight, $w_i$
associated with each index $\mu_i$. The
operators are to be normal-ordered with
respect to the free creation and
annihilation operators, a fact which will
be denoted, when necessary, by the
usual normal ordering symbol, ${:}\;{:}\,$.
In section 1, we gave an abstract
specification of these elements in terms of
tensors and their associated
weights. The origin of the identifications
in equations
(\eqlinear)--(\equnordered) should now be clear:
the operators $h$ are linear
in the tensors for fixed weights, which explains
equation (\eqlinear); normal
ordering means that the order of the weighted
indices does not matter, equation
(\equnordered); equation (\eqtotder) is
simply the statement that the integral
of a total derivative vanishes ($\Delta_r$
differentiates the first $r$ factors
in $h$).

The operators of equation (\eqkorad) are
associated with a {\it deformation
class\/} of CFTs, not with a particular
field theory, and their equal-time
commutators are similarly independent
of any particular choice of CFT. We may
therefore compute the structure constants
in any convenient theory and yet know
that that the result will hold for the
entire deformation class. That is, the
structure constants will apply to string
{\it theory\/}, not just to some
particular solution. It is clearly simplest
to calculate in free field theory,
where the integrands are holomorphic,
and we may easily use the operator
product expansion to compute commutators.

We want to calculate the equal-time commutators
$$
[h_1, h_2]=[\ints {\psi_{\mu\nu\cdots\rho}}{\partial^{w_1}}X^{\mu} \cdots
{\partial^{w_n}}X^{\rho},\int d{\sigma'}
{\chi_{\kappa\lambda\cdots\sigma}}{\partial^{v_1}}X^{\kappa} \cdots
{\partial^{v_m}}X^{\sigma}]\numbereq\name\eqkatou
$$
Our starting point will be the operator product expansion on the complex plane
$$
{\psi_{\mu\nu\cdots\rho}}{\partial^{w_1}}X^{\mu}{\partial^{w_2}}
X^{\nu} \cdots {\partial^{w_n}}X^{\rho}(u) \>
{\chi_{\kappa\lambda\cdots\sigma}}{\partial^{v_1}}X^{\kappa}{\partial^{v_2}}
X^{\lambda} \cdots {\partial^{v_m}}X^{\sigma}(z)\numbereq\name\eqairin
$$
In general this operator product expansion
will include singular terms of the
form ${{{\Theta_m}(w,\overline w)} \over (u-z)^m}$.
Performing the contour
integration and transforming back on
the cylinder will produce terms
${{\delta}^{(m-1)}}({\sigma}-{\sigma'}){\Theta_m}({\sigma'})$.
Since we are
interested
in the commutators of the zero modes of
these operators, most of these terms
will give zero upon
integration over $\sigma$. The only
non-zero result will arise
from simple poles in the operator product
expansion, and it is therefore
the coefficient of this simple pole that we must calculate.

Taking into account that the two-point function for free bosons is
$\left\langle X^\mu(z)X_\nu(w)\right\rangle =-{\log (z-w)}
{\delta^{\mu}_{\nu}}$,
we can prove the following formula by induction:
$$
{\partial^{w_i}}X^{\mu}(u){\partial^{v_j}}X_{\nu}(z)=
	{(-1)^{w_i}({w_i}+{v_j}-1)!
	\over (u-z)^{w_i+v_j}} + \ldots \numbereq\name\eqrut
$$
The problem therefore reduces to a calculation of the simple pole in the
operator product of the two integrands. This is not hard, the principle
difficulty being figuring out how to write Wick's theorem in a way that
sufficiently automates the calculation. This may be done as follows:
$$
\eqalignno{
{:}\prod_{i\in\cal S} \partial^{w_i}X^{\mu_i}(u){:}\>
	{:}\prod_{j\in\cal T} \partial^{v_j}X^{\nu_j}(z){:}
= \sum_{\scriptstyle\cal U\subseteq S\atop\scriptstyle \mathstrut P\colon
U\hookrightarrow T} &\prod_{k\in\cal U}
\left\langle \partial^{w_k}X^{\mu_k}(u)
\partial^{v_{{\cal P}(k)}}X^{\nu_{{\cal P}(k)}}(z)
\right\rangle\cr
&\qquad\qquad{:}\prod_{\hbox to 15pt{\hss$\scriptstyle i\in\cal
	S-U$\hss}} \partial^{w_i}X^{\mu_i}(u)\;
	\prod_{\hbox to 15pt{\hss $\scriptstyle j\in\cal T-P(U)$\hss\hss}}
	\partial^{v_j}X^{\nu_j}(z){:}\>.
&\eqalinno\name\eqwick\cr}
$$
As discussed in section 1, $\cal S$ and
$\cal T$ are index sets labeling the
factors in the two normal-ordered operators
whose product is to be taken. The
sum is over all pairs of subsets $\cal U\in S$ and injections
$\cal P {:} U \hookrightarrow T$,
which is to say, the sum over all
contractions. We now substitute
our formula for the contractions,
equation~(\eqrut), into equation~(\eqwick) to obtain,
$$
\eqalignno{
{:}\prod_{i\in\cal S} \partial^{w_i}X^{\mu_i}(u){:}\>
	{:}\prod_{j\in\cal T} \partial^{v_j}X^{\nu_j}(z){:}
= \sum_{\scriptstyle\cal U\subseteq S\atop\scriptstyle \mathstrut P\colon
U\hookrightarrow T} &\prod_{k\in\cal U} {\eta^{\mu_k\nu_{{\cal P}(k)}}
(-1)^{w_k}(w_k + v_{{\cal P}(k)}-1)!\over (u-z)^{w_k + v_{{\cal P}(k)}}}\cr
&\qquad\qquad{:}\prod_{\hbox to 15pt{\hss$\scriptstyle i\in\cal
	 S-U$\hss}} \partial^{w_i}X^{\mu_i}(u)\;
	\prod_{\hbox to 15pt{\hss $\scriptstyle j\in\cal T-P(U)$\hss\hss}}
	\partial^{v_j}X^{\nu_j}(z){:}\>.
&\eqalinno\name\eqnoprod\cr}
$$
The next step is to extract the simple
pole from equation (\eqnoprod). To do
this we must Taylor expand the $u$-dependent part of the normal-ordered
operators about $z$:
$$
f(u) = \sum_{n=0}^\infty {1\over n!}\partial^nf(z) (u-z)^n,
\numbereq\name\eqtaylor
$$
and observe that, for a given pair $\cal U$
and $\cal P$, the leading pole in
equation (\eqnoprod) is of order $\sum_{k\in\cal U} w_k + v_{{\cal P}(k)}$.
To pick out the simple pole, we must therefore take the term with
$n=\sum_{k\in\cal U} (w_k + v_{{\cal P}(k)}) -1$ from the Taylor expansion,
equation (\eqtaylor). This yields
$$
\eqalignno{
{:}\prod_{i\in\cal S} \partial^{w_i}X^{\mu_i}(u){:} \>
	{:}\prod_{j\in\cal T} \partial^{v_j}X^{\nu_j}(z){:}
= &\sum_{\scriptstyle\cal U\subseteq S\atop\scriptstyle \mathstrut P\colon
U\hookrightarrow T} {\prod_{k\in\cal U} \eta^{\mu_k\nu_{{\cal P}(k)}}
(-1)^{w_k}(w_k + v_{{\cal P}(k)}-1)!\over (u-z) (\sum_{k\in\cal U} (w_k +
v_{{\cal P}(k)}) -1)!}\cr
&\qquad{:} \partial^{\left(\sum_{k\in\cal U}
	(w_k + v_{{\cal P}(k)}) -1\right)}
	\bigl(\>\prod_{\hbox to 15pt{\hss$\scriptstyle i\in\cal S-U$\hss}}
	\partial^{w_i}X^{\mu_i}\bigr)
	\prod_{\hbox to 15pt{\hss $\scriptstyle j\in\cal T-P(U)$\hss\hss}}
	\partial^{v_j}X^{\nu_j}(z){:} + \ldots
&\eqalinno\name\eqsimpole\cr}
$$
where the ellipsis refers to other powers of
$(u-z)$ which are of no interest to us.

Upon contracting both sides of equation (\eqsimpole) with the coefficient
tensors $\psi_{\mu_{\cal S}}$ and
$\chi_{\nu_{\cal T}}$ (here, the subscript
$\mu_{\cal S}$ indicates the full set of indices indexed by $\cal S$), we
read off the commutator,
$$
\eqalignno{
[h_1,h_2] = \sum_{\scriptstyle\cal U\subseteq S\atop\scriptstyle
	\mathstrut P\colon U\hookrightarrow T}\int\,d\sigma\,&
\psi_{\mu_{\cal S}} \chi_{\nu_{\cal T}}
{\prod_{k\in\cal U} \eta^{\mu_k\nu_{{\cal P}(k)}}
(-1)^{w_k}(w_k + v_{{\cal P}(k)}-1)!\over (\sum_{k\in\cal U} (w_k +
v_{{\cal P}(k)}) -1)!}\cr
&\qquad\qquad{:} \partial^{\left(\sum_{k\in\cal U}
	(w_k + v_{{\cal P}(k)}) -1\right)}
	\bigl(\>\prod_{\hbox to 15pt{\hss$\scriptstyle i\in\cal S-U$\hss}}
	\partial^{w_i}X^{\mu_i}\bigr)\;
	\prod_{\hbox to 15pt{\hss $\scriptstyle j\in\cal T-P(U)$\hss\hss}}
	\partial^{v_j}X^{\nu_j}\s{:}\>.&\eqalinno\name\eqcomres\cr
}
$$
This is precisely the commutator given, in more abstract form,
in section~1, so our calculation is completed.

\newsection Problems with More General Commutators.

In the last section, we computed explicitly the
commutators of zero-modes
of {\it holomorphic\/} operators constructed out of
derivatives of scalar
fields. We wish to emphasize that we see no problem with
this calculation,
but we limited ourselves to this subalgebra precisely
because there are
problems associated with more general commutators. This
section is devoted
to a description of these difficulties, and as such its
only logical
connection to its predecessor is that it explains why we
restricted
the scope of our work as we did.

We were able to study the subalgebra of the previous
section because the
fields in question are holomorphic and their operator
products contained
only poles. Unfortunately, this is not always, or even
usually, the case
in string theory. When fields contain functions of a
scalar field (not
just derivatives) the short distance singularities may be
more complicated
since
$$
e^{ip\cdot X_L(z)} e^{iq\cdot X_L(w)} \sim {e^{i(p+q)\cdot
X_L(w)}\over
\left(z-w\right)^{-p\cdot q}},\numbereq\name{\eqnopole}
$$
and $p\cdot q$ need not be an integer.

Moreover, we are {\it not\/} just interested in the
commutators of
(anti-)holomorphic fields; any operator containing both
$\partial^wX$
and $\ov{\partial^v}X$ as factors will not be holomorphic,
and the same
is true for any non-constant function of $X=X_L(z)+X_R(\ov
z)$.
This is important because the relationship between the
operator product
expansion and equal-time commutators holds only if one of
the fields
is (anti-)holomorphic. (Recall that demonstrating this
relationship
involves deforming contours and invoking Cauchy's
theorem.)

Nor are these difficulties easily circumvented. For
example, one might try
a, ``conformal block," representation of a non-holomorphic
operator as
a sum of products of holomorphic and anti-holomorphic
fields, and then
try to construct the full commutator out of the
commutators for the
(anti-)holomorphic components. Unfortunately, a simple
example shows that
this approach is inadequate. Consider the commutator
$\left[A\ov A\s,
B\ov B\sp\right]$, where $A$ and $B$ are holomorphic and
$\ov A$ and
$\ov B$ are anti-holomorphic. Then,
$$
\left[A\ov A\s, B\ov B\sp\right] = \left[A\s,
B\sp\right]\ov B\sp
	\ov A\s + A\s B\sp \left[\ov A\s, \ov B\sp\right].
		\numbereq\name{\eqprob}
$$
Here we have assumed that, as is usually the case, the
holomorphic and
anti-holomorphic operators are constructed from disjoint,
mutually commuting
sets of creation and annihilation operators.

The problem with equation (\eqprob) is that each
commutator will typically
be very local, being constructed out of $\d$ and its
derivatives.
Unfortunately this means that we must take the product of
the fields
outside the commutator ({\it e.g.} $\ov B\sp\ov A\s$ in
the first term)
at the same point---a clear signal of possible trouble.
Note that the
product of fields outside the commutators is {\it not\/}
normal-ordered.
At the very least, then, there is a problem to be
addressed over and
above knowing the  commutators of the (anti-)holomorphic
blocks.

Evidently, we need to calculate commutators with greater
care. We shall
therefore proceed by defining composite fields through a
point-splitting
regularization and renormalization, calculating the
commutators at finite
splitting and finally taking the splitting to zero. This
rather long-winded
method yields the correct result when applied to
holomorphic operators,
and so we may be reasonably confident of its validity.
However, we shall
see that in many cases of interest, taking the
point-splitting to zero
does not yield a finite limit.

It will be instructive to sketch first a calculation that
{\it does\/}
work, and the obvious choice is the Virasoro algebra of
the stress tensor
of a single free scalar, which we define by point
splitting as follows:
$$
\eqalignno{
T\s &= {\textstyle {1\over 2}}{:}\dx\dx\s{:}\cr
	&= \lim_{\epsilon\rightarrow 0} {\textstyle
{1\over 2}}
		\dx\s\dx(\sigma+\epsilon) + {1\over
4\pi\epsilon^2}
		&\eqalinno\name{\eqpointsplit}
}
$$
Hence we may calculate the commutator:
$$
\eqalignno{
4\left[T\s, T\sp\right] &=
\lim_{\epsilon,\epsilon'\rightarrow 0}
\left[\dx\s\dx(\sigma+\epsilon),\dx\sp\dx(\sigma'+\epsilon'
)\right]\cr
\noalign{\vskip 1\jot}
&= \lim_{\epsilon,\epsilon'\rightarrow 0}
i\left(\dx\se\dx\sep\dpr
	+\dx\s\dx\sep\delta'(\sigma+\epsilon-\sigma')\right.\cr
&\qquad\qquad\qquad\left.
+\dx\sp\dx\s\delta'(\sigma+\epsilon-\sigma'-\epsilon')
	+\dx\sp\dx\se\delta'(\sigma-\sigma'-\epsilon')
\right)\cr
\noalign{\vskip 1\jot}
&=4i{:}\dx\s\dx\sp{:} \dpr -
\lim_{\epsilon,\epsilon'\rightarrow 0}
	{i\over 2\pi}
	\left\{
{\dpr\over(\sigma+\epsilon-\sigma'-\epsilon')^2}\right.\cr
&\left.\qquad\qquad\qquad +
{\delta'(\sigma+\epsilon-\sigma')\over
	(\sigma-\sigma'-\epsilon')^2}
	+{\delta'(\sigma+\epsilon-\sigma'-\epsilon')\over
(\sigma-\sigma')^2}
	+{\delta'(\sigma-\sigma'-\epsilon')\over
(\sigma+\epsilon-\sigma')^2}
\right\}&\eqalinno\name{\eqpartway}\cr
}
$$
Now using $f\s\dpr = f\sp\dpr - f'\sp\d$, equation
(\eqpartway) becomes
$$
\eqalignno{
\left[T\s, T\sp\right] &= 2iT\sp\dpr - iT'\sp\d
	-\lim_{\epsilon,\epsilon'\rightarrow 0} {i\over
8\pi}
	\left\{{\dpr+\delta'(\sigma+\epsilon-\sigma'-\epsilon')
		\over (\epsilon-\epsilon')^2} \right.\cr
&\qquad\qquad+{\delta'(\sigma+\epsilon-\sigma') +
\delta'(\sigma-\sigma'
	-\epsilon')\over (\epsilon + \epsilon')^2} +
	2{\d - \delta(\sigma+\epsilon-\sigma'-\epsilon')
		\over (\epsilon-\epsilon')^3}\cr
&\qquad\qquad\qquad\qquad\left.
	+2{\delta(\sigma-\sigma'-\epsilon') -
\delta(\sigma+\epsilon-\sigma')
		\over (\epsilon + \epsilon')^3}
\right\}&\eqalinno\name{\eqclose}\cr}
$$
It is a very non-trivial fact that the right-hand side of
equation
(\eqclose) does have a sensible limit, the non-trivial
term being
${-i\over 24\pi}\dpppr$, in agreement with equation
(\eqvir).
(Of course, to show this properly we should convolve these
distributions
with suitable test functions, but our slightly heuristic
arguments
wherein we, ``differentiate," delta-functions carry over
directly
to a more pedantic proof.)
It is precisely the fact that these singular terms do not
always group
themselves
into nice derivatives of delta-functions that is the
problem we wish
to draw attention to in this section.

Let us therefore now exhibit an example where the above
method fails;
consider the commutator ${\cal C}=\left[e^{ip\cdot
X}\dx\s,\, e^{iq\cdot X}\dx\sp
\right]$. As in equation (\eqpointsplit), we must first
define what we mean by the normal-ordered operators in
this commutator:
$$
{:}e^{ip\cdot X} \dx\s {:} = \lim_{\epsilon\rightarrow
0}\left\{
{:}e^{ip\cdot X\s}{:}\, \dx\se - {i\sqrt2 p\,
{:}e^{ip\cdot X\s}{:} \over
4\pi\epsilon}\right\}.\numbereq\name{\eqsplittwo}
$$
Then the commutator we wish to calculate is
$$
\eqalignno{
{\cal C}=\biggl[e^{ip\cdot X}\dx\s\,,\, e^{iq\cdot
X}\dx\sp\biggr] =
\lim_{\epsilon, \epsilon'\rightarrow 0} \biggl[
{:}e^{ip\cdot X\s}{:}\,\dx&\se - {i\sqrt2 p {:}e^{ip\cdot
X\s}{:} \over 4\pi\epsilon}\,,\cr
&{:}e^{iq\cdot X\sp}{:}\,\dx\sep - {i\sqrt2 q\,
{:}e^{iq\cdot X\sp}{:} \over 4\pi\epsilon'} \biggr].
&\eqalinno\name{\eqprobcom}\cr
}
$$
With a little algebra this expands to
$$
\eqalignno{
{\cal C} = &\lim_{\epsilon, \epsilon'\rightarrow 0}\bigl\{
(q/\sqrt2)\de\, {:}e^{ip\cdot X\spme}{:}\,{:}e^{iq\cdot
X\sp}{:}\,\dx\sep\cr
&\qquad{} - (p/\sqrt2)\dep\, {:}e^{iq\cdot
X\sp}{:}\,{:}e^{ip\cdot X\sep}{:}\,\dx\spaeaep\cr
&\qquad\qquad{} + i\dpreep\,{:}e^{ip\cdot
X\spaepme}{:}\,{:}e^{iq\cdot X\sp}{:}
+ p\deep\,{:}e^{ip\cdot X\spaepme}{:}\,{:}e^{iq\cdot
X\sp}{:}\cr
&\qquad\qquad\qquad{}+
(ip^2/4\pi\epsilon)\dep\,{:}e^{iq\cdot
X\sp}{:}\,{:}e^{ip\cdot X\sep}{:}\cr
&\qquad\qquad\qquad\qquad {}-
(iq^2/4\pi\epsilon')\de\,{:}e^{ip\cdot
X\spme}{:}\,{:}e^{iq\cdot X\sp}{:}\bigr\}&
\eqalinno\name{\eqnonord}\cr
}
$$

As in the case of the Virasoro algebra (the last step in
equation (\eqpartway)), we must now rewrite the right hand
side of equation (\eqnonord) in terms of normal-ordered
operators, and then attempt to take the limits.
In order to do this we need one more piece of
information---how to normal-order products of exponentials
at nearby points. The answer is,
$$
{:}e^{ip\cdot X\s}{:}\,{:}e^{iq\cdot X\se}{:} =
{:}e^{ip\cdot X\s + iq\cdot X\se}{:}\, \epsilon^{p\cdot
q/2\pi},\numbereq\name{\eqnormexp}
$$
where we assume that $\epsilon > 0$. Of course $p\cdot q$
can take any real value, so that in normal ordering these
exponentials (and so in normal ordering any product of
functions of $X$) we may get arbitrarily disgusting
singularities. It is the appearance of these non-integer
powers of $\epsilon$ that, more than anything else, dooms
our attempt to make sense of this operator algebra, since,
in contrast to the Virasoro case above, they are clearly
{\it not\/} going to group themselves into derivatives of
delta-functions. Nevertheless, let us carry the
calculation through to the bitter end. Expressing equation
(\eqnonord) in terms of normal-ordered operators yields
$$
\eqalignno{
{\cal C} = \lim_{\epsilon, \epsilon'\rightarrow 0}\bigl\{
&(-p/\sqrt2){\epsilon'}^{p\cdot q/2\pi}\dep\,{:}e^{iq\cdot
X\sp+ip\cdot X\sep}
\dx\spaeaep{:}\cr
&\qquad{}+(q/\sqrt2)\epsilon^{p\cdot
q/2\pi}\de\,{:}e^{iq\cdot X\sp+ip\cdot
X\spme}\dx\sep{:}\cr
&\qquad\qquad{}+{|\epsilon-\epsilon'|}^{p\cdot
q/2\pi}\left\{i\dpreep
+p\deep\right\}\,{:}e^{iq\cdot X\sp+ip\cdot
X\spaepme}{:}\cr
&\qquad\qquad\qquad{}-{ip\cdot q\over
4\pi(\epsilon+\epsilon')}\bigl(
{\epsilon'}^{p\cdot q/2\pi}\dep\,{:}e^{iq\cdot
X\sp+ip\cdot X\sep}{:}\cr
&\qquad\qquad\qquad\qquad{}- {\epsilon}^{p\cdot
q/2\pi}\de\,{:}e^{iq\cdot X\sp+ip\cdot X\spme}{:}
\bigr)
\bigr\}.&\eqalinno\name{\eqsadresult}
}
$$
Clearly, equation (\eqsadresult) has no sensible limit as
$\epsilon, \epsilon'\rightarrow 0$ for generic $p\cdot
q<0$, and we have failed in our attempt to define the
commutator of equation (\eqprobcom).

We have shown, then, that not all {\it normal-ordered\/} fields
have well-defined commutators. Perhaps this is surprising
at first sight, since we might as well have been dealing
with free-field theory, where we normally think of
normal-ordering as being sufficient. However, a little
thought convinces us that this need not be the case. Any
calculation in an interacting field theory may be reduced
to a calculation in free-field theory with composite
operators. The reason is simply that, in calculating some
Greens function (for example), operators at different
times may be represented as operators at some fixed time
acted on by exponentials of the Hamiltonian (a composite
operator). However, we know full well that there are
divergences in such objects, and that normal ordering is
not a sufficient palliative---there are additional,
logarithmic divergences which need counterterms at each
order in perturbation theory. Indeed, the divergences that
gave rise to the difficulties in equation (\eqsadresult)
are precisely these logarithmic divergences.
${\epsilon}^{p\cdot q/2\pi}$ may not look like
$\ln\epsilon$, but that's only because powers of
logarithms have been summed; if we expand $e^{ip\cdot
X\s}$ we will have to deal with contractions of powers of
$X\s$, which yield logarithms.

It might be thought that restricting our attention to
$(1,1)$ primary fields will save us, since they are
associated with conformal deformations of the stress
tensor, and conformal field theories are ultra-violet
finite. Alas, this is not so; in general, $(1,1)$
primaries preserve conformal invariance only to {\it
first\/} order in the deformation, and deforming by a {\it
finite\/} (1,1) primary will, in general, yield a
non-conformal theory which needs further renormalization.
(Deformations where this does not occur are sometimes
called {\it strictly marginal}.)

This concludes our arguments in support of the proposition
that the {\it normal ordered\/} operators encountered in
string theory are frequently sick. However, even the
astute reader who accepts these arguments may nevertheless
argue that this problem is easily fixed. Surely it is
sufficient to add the appropriate logarithmic counterterms
to the definitions of our operators. That is, we should
add terms involving $\ln \epsilon$ to the right hand sides
of equations (\eqnormexp), (\eqsplittwo) and, perhaps,
(\eqpointsplit). Alas! we shall argue that our experience
with renormalizable field theories indicates that such a
strategy need not succeed.

The reason is that the task we have set ourselves is more
demanding than the one we usually face when renormalizing
a field theory. In that case we need only make sense of
{\it one\/} composite operator, namely the Hamiltonian.
We, on the other hand, must make sense of an {\it
infinite\/} number of composite operators, assigning fixed
counterterms to each in such a way that the commutator of
any pair is well defined. Consider equation
(\eqsadresult), the result of calculating the commutator
in equation (\eqprobcom). To the first operator in
(\eqprobcom) we must add counterterms {\it depending only
on\/} $p$, while the corresponding counterterms with
$p\rightarrow q$ should be added to the second operator.
However, it is very hard to imagine how such counterterms
could cancel the $\epsilon^{p\cdot q/2\pi}$ terms in
equation (\eqsadresult).

There is a simpler and more compelling example of this
difficulty in renormalizable field theory.\footnote*{This
argument was developed in conversation with Alex Kovner.}
In that case, the renormalization procedure makes sense of
an interacting Hamiltonian, $H$, that is formally written
as the sum of a free part and an interacting part,
$$
H=H_0+gH_I,  \numbereq\name\eqsplitham
$$
where $g$ is the coupling. However, we shall argue that
$H_0$ and $gH_I$ are {\it not\/} in fact {\it
separately\/} well defined. The reason is that $g$ depends
on the regulator, $\epsilon$; typically,
$g(\epsilon)=1/(1+\beta\ln (\mu\epsilon))$, where $\mu$ is
some arbitrary renormalization scale and $\beta$ is some
calculable number. The {\it only\/} freedom we have in
defining $H$ is in the variation of $\mu$. On the other
hand, if $H_0$ and $gH_I$ were separately well defined, we
would be able to add them with an arbitrary relative
weight: $H=H_0+\lambda gH_I$, with $\lambda$ some finite
($\epsilon$-independent) number. This, however, does {\it
not\/} correspond to a finite ($\epsilon$-independent)
rescaling of $\mu$ (it requires an unacceptable change in
$\beta$ also, to put it back in canonical form). We
therefore conclude, {\it reductio ad absurdum}, that $H_0$
and $gH_I$ are {\it not\/} simultaneously well-defined
(although their sum is). Thus the problem of
simultaneously defining several composite operators is
not, in general, solved by the conventional
renormalization procedure.

As we said earlier, we are not sure what to make of this
observation. It would be premature to conclude that there
is something wrong with string theory, since we seem to be
able to construct perfectly satisfactory scattering
amplitudes. However, it does suggest that arguments based
on the existence of an algebra of composite operators need
to be treated with caution. Perhaps the way round this
difficulty is to use the operator formalism \ref{\oper}
{L. Alvarez-Gaum\'e, C. Gomez, G. Moore and C. Vafa \np303
(1988), 455;
M. Campbell, P. Nelson and E. Wong \imp6 (1991), 4909.},
where,
despite its name, the operator algebra is superfluous.
Indeed, there are proposals for how to make finite
deformations of conformal field theories within that
formalism \ref{\findef}{K. Ranganathan, H. Sonoda and
B. Zwiebach preprint MIT-CTP-2193, hep-th/9304053; G. Pelts
Rockefeller University preprint RU-93-5-B, hep-th/9309141 and
preprint in preparation.}, a problem that would seem to require
solving or evading the difficulty pointed out here.

\newsection Conclusions.

Let us review the results presented
in this paper. We gave a rather sketchy
outline of previous work on deformations
of conformal field theories and
its relation to the symmetries of string
theory. For the purposes of this
paper, the important idea to be drawn
from this work is that symmetries are
generated by a subalgebra of the full
operator algebra of (the deformation
class of) a conformal field theory. The
subalgebra in question is the
centralizer of the generators of world-sheet
translations, which is a fancy
way of saying all zero modes of quantum
fields. It should also be noted that
this statement may only be true after
the introduction of auxiliary
space-time fields, as in supersymmetric field theories.

However, most of the operators of
interest are composite; they involve
products of elementary fields at the
same point, and so are potentially
sick. We showed that such fears are
justified; many of the normal-ordered
composite operators that we habitually
use in string theory (including,
for example, vertices and currents)
do not have well defined commutators.

Nevertheless, we did identify an
infinite set of fields that do have
well-defined commutators between themselves,
and we computed the commutators
of an arbitrary pair of zero modes. This then constitutes
an infinite-dimensional subalgebra of the
full symmetry of string theory.

This algebra is unfamiliar to us, and
differs from most of the infinite
algebras that have appeared in physics
in that the infinite-dimensionality
does not
arise through the moding of a finite
number of fields, but rather through
having an infinite number of fields
in the first place. The closest
relative of our algebra would appear
to be $W_\infty$ \ref\pope{C.\tie
Pope, L.\tie Romans and X.\tie Shen,
\pl236 (1990), 173.}, which may
be realized in terms of the fields of
equation (\eqkorad) with weights
$w_i = 1$ \ref\hull{C.\tie Hull, \np353
(1991), 707.}. However, to obtain
$W_\infty$ it is necessary to retain
{\it all\/} modes of these fields,
while the zero-modes form only the Cartan
subalgebra. The apparent kinship
is therefore rather distant.

Our algebra
naturally possesses the general properties
of string symmetries remarked
in earlier work [\evao], [\egw]: it is a
supersymmetry in that its
generators do not commute with the generators
of Lorentz transformations and
mix excitations of different spin; it is
spontaneously broken in flat
space-time because not all generators
commute with the stress tensor
of the free scalar CFT, and it transforms
excitations of differing mass
into one another.

Naively, the symmetries we are discussing
should be local symmetries. If the
constant tensors ${\psi}$ and $\chi$ in the
generators were to depend
upon the scalar field, $X^{\mu}(\sigma)$,
the locality would be apparent, but
it is precisely the appearance of $X$-dependence
that wrought havoc with
commutators, as we saw in section~4. In some
sense, then, our sub-algebra
would appear to be the {\it global\/} part
of a gauge symmetry. This is rather
encouraging, since global parts are usually
the repository of most physics
information (relations between couplings,
conserved charges, {\it etc}.).
However, the full global algebra should
presumably include both holomorphic
and anti-holomorphic derivatives, and
prior work suggests that transformations
involving {\it propagating\/} (as opposed
to auxiliary) degrees of freedom
would be generated by
operators more evenly balanced between the
two types of derivatives. Thus
the physical importance of our sub-algebra
remains to be determined.
Commutators between such mixed generators
are not calculable by the
methods of section 3, but may yield those of
section 4. This question is
currently under investigation.

As we have remarked, the pathologies associated
with the various composite
operators of string theory are something
of an enigma. Despite this problem,
they seem to permit us to calculate perfectly
fine scattering amplitudes.
We can only hope that further work will shed
some light on this puzzle.

\immediate\closeout1
\bigbreak\bigskip

\line{\twelvebf References. \hfil}
\nobreak\medskip\vskip\parskip

\input refs

\vfil\end